\documentclass[letterpaper,11pt,leqno]{article}
\usepackage{paper}

\hypersetup{pdftitle={Incentives to Offer Algorithmic Recourse}}

\begin{document}

\title{Incentives to Offer Algorithmic Recourse}

\author{Matthew Olckers 
\and Toby Walsh
%
\thanks{Olckers: Centre for Household Finance, Stellenbosch University. Walsh: UNSW Sydney. We thank Francis Bloch for helpful comments.}}

\date{September 2026}   


\begin{titlepage}
\maketitle
Algorithmic recourse promises to help applicants rejected by automated systems by explaining the changes needed to secure acceptance. What incentive do decision-makers, such as banks and employers, have to offer recourse? We study this question in a screening model in which recourse is both productive and selective: completing recourse improves an applicant's value to the decision-maker, but applicants differ in their cost of completion. The optimal policy is a threshold rule: reject applicants with low scores, offer recourse to an intermediate range of scores, and accept applicants with high scores outright. Because the intermediate range spans the cutoff that would separate acceptance from rejection when recourse is not available, some marginal applicants gain a new path to acceptance, while others---who would have been accepted outright---must now clear a costly hurdle.
\end{titlepage}

\section{Introduction}

Algorithms decide whether applications are accepted or rejected in many high-stakes domains such as credit, hiring, insurance, and social services. Applicants receive instant decisions but rarely receive any explanation. ``Your application was rejected. Why? Because our algorithm said so.'' How should rejected applicants respond? One solution is \emph{algorithmic recourse}: when an individual is rejected, the system explains the concrete changes that would lead to a favorable decision.

Apple's credit card provides a clear example of algorithmic recourse. Applicants who are rejected may be invited to join the ``Path to Apple Card''. The program sets personalized steps, such as making required payments on time, reducing credit card or personal loan debt, or resolving past-due balances.\footnote{See Apple's description of Path to Apple Card at \url{https://support.apple.com/en-us/109298}. Apple reported in January 2024 that more than 200,000 users had been approved for Apple Card after enrolling in and completing the program; see \url{https://www.apple.com/newsroom/2024/01/apple-card-is-helping-cardholders-live-healthier-financial-lives/}.} As one applicant explains, their Path to the Apple Card was to ``pay off one closed credit card with a balance and reported on collections, and lower my credit card debt from over \$4000 to under \$670 in 4 months.''\footnote{See the full applicant comment and similar discussions at \url{https://www.reddit.com/r/AppleCard/comments/1sdut5z/comment/oepa1xw/}}

Yet Path to Apple Card is not offered to all rejected applicants, in conflict with the normative case for recourse. Writing at the intersection of algorithmic fairness and moral philosophy, \citet{venkatasubramanian2020} argue that recourse is valuable because it supports trust and the ability to make long-term plans. Many long-term goals require a sequence of smaller decisions made by algorithms or bureaucracies (e.g., obtaining a loan to buy a car to get to work). Recourse is valuable not merely because it reverses a single unfavorable decision, but because it gives people reason to expect that, were something to go wrong, they would be able to set it right. The argument implies recourse should be available to everyone.

We analyze the incentives for a principal, such as a bank or employer, to offer recourse. We study this question in a simple screening model. The principal faces an applicant who may be profitable or unprofitable but observes only a noisy score, such as a machine-learning prediction. The principal can reject the applicant, accept him, or offer recourse. If recourse is offered, the principal chooses a requirement that determines how demanding the recourse process is (for example, how much the applicant must increase his income or reduce his debt). Applicants differ in the cost of completing recourse, and completion raises the principal's payoff by improving the applicant's effective quality. The principal thus uses recourse both to increase applicants' value and to screen them. More profitable applicants have weakly lower expected recourse costs and are therefore at least as likely to complete the process.

Our main result is that recourse can harm applicants. Relative to a benchmark in which recourse is unavailable and the principal can only accept or reject, making recourse available can cause the principal to impose a costly requirement on applicants who would otherwise be accepted outright. In the case of Path to Apple Card, this additional hurdle can require months of timely payments and debt reduction before an applicant can receive their credit card. We develop our main result in the following steps.

Our first result characterizes when recourse replaces rejection over the full posterior interval. Either a requirement is sufficiently productive to offset the loss from accepting an unprofitable applicant, or a requirement is sufficiently demanding that unprofitable applicants never complete it.

We then study the case in which feasible recourse is limited: it improves applicants but does not fully compensate for the loss from low types, and it is not stringent enough to screen low types out completely. In this case, recourse is valuable in an intermediate range. An applicant is not sufficiently attractive to accept unconditionally but also not sufficiently unattractive to reject. The optimal policy is a threshold rule: reject applicants with low scores, offer recourse to applicants with intermediate scores, and accept applicants with high scores.

With this threshold, we compare the welfare of applicants under recourse to their welfare in a benchmark where recourse is unavailable and the principal accepts or rejects based on a single cutoff score. Recourse replaces the single cutoff with a range of intermediate signals where the principal offers recourse instead of an outright accept or reject. Applicants at the bottom of the range are better off, since they would have been rejected in the benchmark and now have a chance at acceptance. Applicants at the top are worse off, since they would have been accepted outright in the benchmark and must now complete a costly requirement to reach the same outcome.

Finally, we show that making recourse more productive need not make existing recourse offers less stringent. Although higher productivity expands the recourse region, it can induce the principal to raise the requirement and thereby reduce the welfare of applicants already offered recourse.

\paragraph{Related literature} We contribute to the literature on screening by studying a hurdle that is itself productive. In the baseline model of \citet{stiglitz1975screening}, costly screening separates types, but ``there are no increases in production from sorting individuals.'' Ordeal mechanisms and self-targeting similarly use costly requirements to induce selection. \citet{nichols1982targeting} discuss restrictions that screen transfer recipients, while \citet{nichols1971waiting} model waiting time as a deadweight loss. Work requirements could, in principle, combine screening with production. \citet{besley1992workfare}, however, abstract from this channel: ``In order to focus exclusively on incentive arguments for workfare, we will assume that the work done in the public sector is unproductive.'' By contrast, completing recourse in our model both selects applicants and increases their value to the principal. We study how this interaction determines who is offered the hurdle and how demanding it is.

Our model is also related to the literature on testing and probabilistic verification. From the principal's perspective, recourse is a test whose pass rate depends on the stringency of the requirement, while immediate acceptance and rejection are degenerate tests that every applicant passes or fails, respectively. We differ from this literature in two respects. First, we ask a different question. \citet{ball2025} ask when a collection of probabilistic tests can be reduced to a most-discerning test for each reported type and how verification changes the optimal mechanism. \citet{rosar2017test} asks how informative a test should be when greater accuracy may discourage voluntary participation. \citet{hancart2025} asks when a principal benefits from allowing an informed agent to choose among several tests. We instead take the principal's posterior belief as given and ask whether to accept, reject, or offer recourse, and how stringent the recourse requirement should be. Second, tests in this literature reveal or verify information about an applicant's existing type but do not change that type. In our model, completing recourse increases the applicant's value to the principal. A more stringent requirement therefore not only changes which applicants complete recourse but also directly increases the principal's payoff from those who do.

Most of the literature on using algorithmically generated scores to decide whether to accept or reject applicants focuses on manipulation. Before the score is generated, applicants may invest in or manipulate their features to improve the principal's assessment and thereby increase their probability of acceptance \citep{hardt2016strategic,bjorkegren2020manipulation,frankel2019muddled,frankel2022improving,perez2022}. This extends the signaling logic of \citet{spence1973}: applicants take costly actions to influence the principal's beliefs and, ultimately, its acceptance decision. We instead study the downstream decision, after the principal has formed a posterior belief: whether to offer recourse and, if so, how demanding the recourse process should be.

Finally, the growing computer science literature on algorithmic recourse, starting with \citet{ustun2019actionable} and surveyed by \citet{karimi2020survey}, asks how to generate recommendations that an applicant can feasibly act upon. A change that would reverse an algorithm's decision need not constitute meaningful recourse. For example, even if a lower age would raise an applicant's score, an algorithm should not recommend that the applicant decrease their age. More generally, this literature constrains recommendations to respect which features an applicant can change, the costs of changing them, and the relationships among features. It thus takes the provision of recourse as given and studies what form a valid recommendation should take. We instead study the prior question of whether the decision-maker finds it optimal to offer recourse at all.

Two recent papers also study the decision-maker's incentives, but, like the signaling models above, focus primarily on manipulation. \citet{harris2021bayesian} study how a decision-maker can use information about an unknown assessment rule to persuade applicants to take desirable actions. \citet{chen2025} show that a decision-maker may withhold recourse because recommendations reveal information that other applicants can use to manipulate the classifier. Our model has no information spillovers across applicants. Instead, the principal withholds recourse because the offer may be unprofitable given the applicant's posterior, the productivity of the requirement, and the probability that each type completes it.

\section{Model}

\paragraph{Players and actions} A principal evaluates an applicant who is either profitable (type $H$) or unprofitable (type $L$). Let $\theta \in \{H,L\}$ denote the applicant's type.

The principal can take one of three actions: accept the applicant, reject the applicant, or offer recourse. Recourse is a set of additional steps the applicant can undertake to gain acceptance. For example, in the context of a loan application, recourse could specify the additional income the applicant needs to earn to receive a loan.

If recourse is offered, the principal specifies a requirement $r \in \mathcal{R}\equiv[\underline{r},\bar{r}]$, where $0<\underline{r}\le \bar{r}$. The requirement captures the stringency of the recourse process for the applicant. The applicant chooses whether to complete recourse. If the applicant completes the recourse process, the applicant is accepted. If the applicant chooses not to complete the recourse process, the applicant is rejected.

\paragraph{Information} The principal has a prior belief that the applicant is profitable:
\begin{equation*}
\Pr(\theta = H) = \pi \in (0,1), \qquad \Pr(\theta = L) = 1-\pi .
\end{equation*}
Before choosing an action, the principal observes a signal $s \in S \subseteq \mathbb{R}$. Conditional on type $\theta$, the signal has density $f_\theta(s)$, where $f_H$ and $f_L$ share the common support $S$. We assume the signal satisfies the monotone likelihood-ratio property:
\begin{equation*}
\frac{f_H(s)}{f_L(s)} \quad\text{is strictly increasing in } s .
\end{equation*}
Thus higher signals are more favorable evidence that the applicant is profitable.

Let $p(s)\equiv \Pr(\theta=H\mid s)$ denote the principal's posterior belief that the applicant is profitable after observing $s$. By Bayes' rule,
\begin{equation*}
p(s) = \frac{\pi f_H(s)}{\pi f_H(s)+(1-\pi)f_L(s)} .
\end{equation*}
The posterior odds satisfy $\tfrac{p(s)}{1-p(s)} = \tfrac{\pi}{1-\pi}\cdot\tfrac{f_H(s)}{f_L(s)}$, so the monotone likelihood-ratio property implies that $p(s)$ is strictly increasing in $s$. Since the principal's payoffs depend on the signal only through the posterior belief, we write the principal's problem directly in terms of the posterior $p \in [0,1]$.

\paragraph{Payoffs} If the applicant is accepted, he obtains payoff $v>0$. If the applicant is rejected or withdraws from recourse, he obtains payoff $0$. If recourse with requirement $r>0$ is offered and the applicant completes recourse, he receives
\[
v - C(r,\theta,\varepsilon).
\]
The recourse cost depends on the requirement $r$, the applicant's type $\theta$, and an idiosyncratic cost shock $\varepsilon$:
\begin{equation*}
C(r,\theta,\varepsilon)= K_{\theta}(r) + \varepsilon,
\end{equation*}
where
\[
K_\theta(r)\geq 0,\qquad K_\theta'(r)>0,\qquad K_\theta''(r)\geq 0
\quad\text{for all }\theta\in\{H,L\}\text{ and }r\in\mathcal R.
\]
We assume the type-specific cost component $K_\theta(r)$ is weakly lower for the profitable type:
\begin{equation}
K_H(r) \leq K_L(r) \quad\text{for all } r \in \mathcal{R}.
\label{ass:cost_ordering}
\end{equation}
That is, in expectation profitable (type~$H$) applicants find recourse no more costly to complete than unprofitable (type~$L$) applicants. The shock $\varepsilon$ is independent of $\theta$ and $s$ and is uniformly distributed, $\varepsilon \sim \text{Unif}[0,\bar{\varepsilon}]$, with $\bar{\varepsilon}>0$.

An applicant of type $\theta$ completes recourse whenever $v-K_\theta(r)-\varepsilon\ge 0$. Hence the completion probability is
\[
q_\theta(r)
\equiv
\Pr\{C(r,\theta,\varepsilon)\le v\}
=
\min\left\{\max\left\{\frac{v-K_\theta(r)}{\bar \varepsilon},0\right\},1\right\}.
\]
In particular, $q_\theta(r)>0$ if and only if $K_\theta(r)<v$, while $q_\theta(r)=0$ if and only if $K_\theta(r)\ge v$.

If the principal accepts an applicant without recourse, the payoff is $a_H$ for a type-$H$ applicant and $-a_L$ for a type-$L$ applicant, where $a_H>0$ and $a_L>0$. If the application is rejected or the applicant withdraws from recourse, the principal's payoff is zero. If the applicant completes requirement $r$ and is accepted, the principal's payoff is
\[
\begin{cases}
a_H+\delta r, & \theta=H,\\
-a_L+\delta r, & \theta=L,
\end{cases}
\]
where $\delta>0$ measures the incremental value generated by completion of recourse.

To break ties, if the payoff from any two actions is the same, the principal chooses accept over recourse, accept over reject, and recourse over reject.

For any posterior belief $p\in[0,1]$, the principal's expected payoffs from the three kinds of actions are
\begin{align}
\text{Accept:} \quad & U^A(p) = p a_H + (1-p) (-a_L), \label{eq:accept_payoff} \\
\text{Reject:} \quad & U^R = 0, \\
\text{Offer recourse:} \quad & U^{O}(p,r) = p q_H(r) (a_H+ \delta r) + (1-p) q_L(r) (-a_L + \delta r), \label{eq:recourse_payoff}
\end{align}
where $q_\theta(r)$ denotes the probability that a type-$\theta$ applicant completes recourse with requirement $r$. These expressions show that the principal's optimal action depends on the signal only through the posterior belief $p(s)$.

Here $r$ denotes a candidate requirement. The principal's decision rule can assign a different requirement to each posterior, so the requirement selected conditional on offering recourse may vary with $p$. We denote an optimal selection by
\[
r^*(p)\in\arg\max_{r\in\mathcal R}U^O(p,r).
\]

\paragraph{Timing} Let $\mathcal{A}=\{A,R\}\cup(\{O\}\times\mathcal{R})$, where $A$ denotes acceptance, $R$ denotes rejection, and $(O,r)$ denotes offering recourse with requirement $r\in\mathcal{R}$. Because the posterior is sufficient for the principal's payoffs, a decision rule specifies a mapping $\varphi:[0,1]\to\mathcal A$ from posterior beliefs to actions. If recourse is offered at posterior $p$, the rule takes the form $\varphi(p)=(O,r(p))$, where $r(p)\in\mathcal R$ is the posterior-contingent requirement. The action following signal $s$ is $\varphi(p(s))$. The sequence of events is:
\begin{enumerate}
    \item The principal commits to a decision rule $\varphi$.
    \item Nature draws the applicant's type $\theta \in \{L,H\}$ with $\Pr(\theta=H)=\pi$. The applicant observes $\theta$; the principal does not.
    \item Conditional on $\theta$, the signal $s$ is drawn from the density $f_\theta$ and observed by the principal.
    \item The principal applies the decision rule, executing the predetermined action $\varphi(p(s))$. If $\varphi(p(s))=(O,r(p(s)))$, the applicant observes the posterior-contingent requirement $r(p(s))$.
    \item If $\varphi(p(s))=(O,r(p(s)))$, an idiosyncratic cost shock $\varepsilon \sim \text{Unif}[0,\bar{\varepsilon}]$ is realized and observed by the applicant. The applicant chooses whether to complete the recourse requirement $r(p(s))$ or withdraw. If he completes recourse, the application is accepted.
    \item Payoffs are realized according to the applicant and principal payoff functions defined above.
\end{enumerate}

\subsection{Connection to algorithmic recourse}

In typical algorithmic recourse settings, a decision-maker observes a feature vector characterizing an applicant (e.g., income, employment history, credit utilization). A classifier or scoring model then maps this feature vector to a score, which is used to determine acceptance or rejection. In our model, the signal $s\in\mathbb{R}$ plays the role of this algorithmic score: it is a summary statistic derived from the applicant's observable characteristics and is the sole basis for the principal's inference about the applicant's unobserved type $\theta$. By working directly with the score $s$ rather than the underlying feature vector, we abstract from the details of how the score is constructed and focus on the principal's optimal response to the information $s$ conveys.

The recourse action is modeled through the requirement $r \in \mathcal{R}$. In standard recourse settings, a recommendation specifies how much an applicant must change actionable features (e.g., income, debt-to-income ratio, savings) to cross a decision boundary. The scalar $r$ summarizes the stringency or magnitude of these required changes.

A key aspect of algorithmic recourse is that the cost of implementing recommendations is heterogeneous and depends on characteristics of the individual. This heterogeneity is captured by the cost function $C(r,\theta,\varepsilon)=K_\theta(r)+\varepsilon$ where $K_\theta(r)\ge 0$, $K_\theta'(r)>0$, $K_\theta''(r)\ge 0$, and $\varepsilon$ is an idiosyncratic shock.

Unlike purely strategic models where individuals alter observed features without changing underlying fundamentals, algorithmic recourse is typically justified by the idea that recommended actions genuinely improve an applicant's quality \citep{karimi2021facct, karimi2020survey}. This is represented by the principal's incremental value term $\delta r$. When an applicant completes recourse, the principal's payoff increases by $\delta r$ relative to accepting the same applicant without recourse, meaning that the action changes the applicant's effective profitability (e.g., higher income reduces default risk). The parameter $\delta$ measures how valuable recourse-induced changes are.

Finally, the model captures that recourse recommendations are optional. Once recourse is offered, the applicant observes the requirement $r$ and decides whether the benefit of acceptance justifies the cost of completing the recommended changes. The applicant completes recourse only when $v-C(r,\theta,\varepsilon)\geq 0$; otherwise he withdraws and is rejected. Recourse take-up therefore depends both on how demanding the requirement is and on applicant-level cost heterogeneity.

In summary, $\varphi$ captures the algorithmic policy governing when recourse is offered, while $r(p)$ captures how the stringency of the recourse recommendation may vary with the posterior. For any candidate requirement $r$, $C(r,\theta,\varepsilon)$ captures heterogeneity in the cost of completing recourse across applicants, and $\delta r$ captures the extent to which recourse corresponds to genuine improvement. Together these elements provide a compact but faithful abstraction of algorithmic recourse systems.

\section{Optimal decision rule}

One of the aims of algorithmic recourse is to offer recourse to all applicants who would otherwise be rejected if the recourse technology was not available. This universal accessibility is crucial because, as \citet{venkatasubramanian2020} argue, recourse protects people's ability to plan their lives by allowing them to take steps to reverse an adverse outcome. Without mechanisms to reverse adverse outcomes, individuals face uncertainty that undermines their long-term decision-making. Under what conditions will the principal prefer recourse over reject for all applicants?

To answer this question, we identify when recourse is preferred to rejection for any posterior. The following proposition shows that rejection is displaced by recourse if and only if recourse can either compensate for the loss from accepting an unprofitable applicant or screen unprofitable applicants out entirely.

\begin{proposition}[Compensation and screening]
\label{prop:compensation_screening}
The tie-broken optimal action is never rejection at any posterior $p\in[0,1]$ if and only if at least one of the following conditions holds:
\begin{align}
\textup{(C) Compensation:}\qquad
&\exists r\in\mathcal R
\text{ such that }
K_L(r)<v
\text{ and }
r\ge \frac{a_L}{\delta} ,
\label{eq:cond_compensation}\\[4pt]
\textup{(S) Screening:}\qquad
&\exists r\in\mathcal R
\text{ such that }
K_L(r)\ge v .
\label{eq:cond_screening}
\end{align}
\end{proposition}

Proposition~\ref{prop:compensation_screening} identifies two ways recourse can dominate rejection at every posterior. Under \textup{(C)}, there is a feasible requirement whose completion yields a nonnegative principal payoff even for a type-$L$ applicant. Under \textup{(S)}, there is a feasible requirement that no type-$L$ applicant completes. The availability of either offer is sufficient to ensure that rejection is never selected.\footnote{Proposition~\ref{prop:compensation_screening} characterizes the optimal rule on the full posterior interval $[0,1]$. The attainable posterior range may exclude neighborhoods of either endpoint. Rejection can therefore be absent on path even when neither \textup{(C)} nor \textup{(S)} holds.}

Proposition~\ref{prop:compensation_screening} describes the two cases in which rejection disappears entirely. We now consider their complement. Suppose every feasible requirement is completed with positive probability by type-$L$ applicants, yet accepting a type-$L$ applicant who completes any such requirement still yields a strictly negative payoff.

\begin{assumption}[Limited feasible recourse]
\label{ass:limited_recourse}
The maximal feasible recourse requirement is too small to compensate for type-$L$ losses and too small to screen out all type-$L$ applicants:
\[
\bar r<\frac{a_L}{\delta}
\qquad\text{and}\qquad
K_L(\bar r)<v .
\]
\end{assumption}

Assumption~\ref{ass:limited_recourse} rules out both channels in Proposition~\ref{prop:compensation_screening}. The first inequality implies that no feasible requirement compensates for the loss from a type-$L$ applicant, so condition~\textup{(C)} fails. The second inequality and the monotonicity of $K_L$ imply that $K_L(r)<v$ for every feasible $r$, so type-$L$ applicants complete each requirement with positive probability and condition~\textup{(S)} fails. The cost ordering~\eqref{ass:cost_ordering} then ensures that type-$H$ applicants also complete every feasible requirement with positive probability.\footnote{In particular, $q_H(r)$ is bounded below by $\min\{(v-K_H(\bar r))/\bar\varepsilon,1\}>0$.}

Although this assumption implies that rejection remains optimal for sufficiently low posteriors, it does not reveal where recourse is preferred to accept and reject. That comparison is complicated by the principal's ability to choose the requirement as a function of $p$. We therefore summarize the best payoff available from recourse with the \emph{recourse value function}
\begin{equation}
V^O(p)\equiv\max_{r\in\mathcal R}U^O(p,r),
\label{eq:recourse_value_function}
\end{equation}
the highest payoff the principal can obtain from recourse at posterior $p$ once the requirement has been chosen optimally. The optimal action at $p$ is then the largest of the three values $U^A(p)$, $U^R=0$, and $V^O(p)$, and the choice of requirement is absorbed into $V^O$.

Whether the optimal rule contains an acceptance region at all depends on the behavior of recourse at the top of the posterior scale, where the applicant is known to be profitable. Accepting a sure type-$H$ applicant yields $a_H$ with certainty, whereas offering recourse replaces that certain payoff with the conditional payoff $q_H(r)(a_H+\delta r)$: the principal forgoes $a_H$ when the applicant withdraws, but receives the productivity bonus $\delta r$ when he completes. The following condition states that this tradeoff is never worthwhile.

\begin{assumption}[Acceptance dominance at the top]
\label{ass:endpoint}
For every feasible requirement $r\in\mathcal R$,
\[
q_H(r)\,(a_H+\delta r)\le a_H .
\]
\end{assumption}

Equivalently, $(1-q_H(r))a_H\ge q_H(r)\delta r$: the expected loss from a profitable applicant failing to complete recourse is at least as large as the productivity gain from completion. Assumption~\ref{ass:endpoint} therefore ensures that acceptance is optimal when the applicant is known to be profitable. Because the inequality is weak, the assumption permits an acceptance region consisting only of $p=1$; acceptance on an interval below $1$ requires strict dominance at the top.\footnote{Proposition~\ref{prop:structure} in the appendix shows that the weak condition is necessary and sufficient for acceptance to be selected somewhere on the full posterior interval. If it fails, recourse strictly dominates outright acceptance at every posterior.}

The two assumptions restrict the principal's choice at the endpoints of the posterior interval: limited feasible recourse makes rejection optimal at $p=0$, while acceptance dominance at the top makes acceptance optimal at $p=1$. To understand the choice between these endpoints, we introduce an interior benchmark that we will use repeatedly: the posterior at which the principal is indifferent between acceptance and rejection. Setting $U^A(p)=0$ gives
\[
p^A\equiv\frac{a_L}{a_H+a_L}\in(0,1).
\]
We call $p^A$ the no-recourse acceptance cutoff. Without recourse, the principal rejects below $p^A$ and accepts at or above it.

At this benchmark, acceptance and rejection both yield zero, whereas recourse yields a strictly positive payoff: completion is at least as likely among type-$H$ applicants as among type-$L$ applicants and generates a productivity bonus. Recourse is therefore optimal at an interior posterior, between the endpoints where rejection and acceptance are optimal. These comparisons yield the following threshold rule.
 
\begin{proposition}[Threshold rule]
\label{prop:threshold}
Suppose Assumptions~\ref{ass:limited_recourse} and~\ref{ass:endpoint} hold. Then there exist cutoffs
\[
0<\underline p<p^A<\bar p\le 1
\]
such that the tie-broken optimal decision rule is
\[
\varphi^*(p)=
\begin{cases}
R, & p\in[0,\underline p),\\[2pt]
(O,r^*(p)), & p\in[\underline p,\bar p),\\[2pt]
A, & p\in[\bar p,1],
\end{cases}
\]
where $r^*(p)\in\arg\max_{r\in\mathcal R}U^O(p,r)$ at every $p\in[\underline p,\bar p)$.
\end{proposition}

Proposition~\ref{prop:threshold} shows why recourse is offered around the no-recourse acceptance cutoff. Below $p^A$, recourse can make a conditional offer profitable where outright acceptance is not. Above $p^A$, recourse can remain more profitable than outright acceptance because only applicants who complete the requirement are accepted and completion generates the additional payoff $\delta r$. For sufficiently low posteriors, type-$L$ losses make rejection optimal; for sufficiently high posteriors, the certain payoff from acceptance dominates recourse. The recourse region therefore reflects the principal's incentives on both sides of the old acceptance cutoff, not only the provision of recourse to applicants who would otherwise be rejected.

The maintained assumptions also imply that completion probabilities are interior for every feasible requirement:
\[
0<q_L(r)\le q_H(r)<1
\qquad\text{for every }r\in\mathcal R.
\]
Limited feasible recourse and the cost ordering give the first two inequalities, while acceptance dominance at the top gives $q_H(r)\le a_H/(a_H+\delta r)<1$.

The threshold rule also yields comparative statics for the recourse region. The cutoffs satisfy
\[
V^O(\underline p)=0,
\qquad
V^O(\bar p)=U^A(\bar p).
\]
To determine how the cutoffs change, we first compare payoffs holding the requirement fixed, and then show that these comparisons carry over to the maximized recourse payoff.

\begin{proposition}[Comparative statics of the recourse region]
\label{prop:comparative_statics}
Suppose Assumptions~\ref{ass:limited_recourse} and~\ref{ass:endpoint} hold and the upper cutoff is interior, $\bar p<1$. Consider changes in one parameter or cost function at a time that preserve these conditions and the remaining assumptions of the model. Then:
\begin{enumerate}[label=\textup{(\alph*)}]
    \item A higher productivity parameter $\delta$ lowers $\underline p$ and raises $\bar p$.
    \item A pointwise weak increase in $K_L$, holding $K_H$ fixed, weakly lowers $\underline p$ and weakly raises $\bar p$. Both movements are strict if $K_L$ increases strictly at every feasible requirement.
    \item A higher upper bound $\bar\varepsilon$ on the idiosyncratic cost shock leaves $\underline p$ unchanged and lowers $\bar p$.
    \item A higher loss $a_L$ raises both $\underline p$ and $\bar p$. A higher gain $a_H$ lowers both cutoffs.
\end{enumerate}
\end{proposition}

A more productive recourse action raises the payoff from every applicant who completes it. Recourse therefore becomes profitable at a lower posterior and remains preferable to unconditional acceptance up to a higher posterior. Greater separation in completion costs works through screening rather than productivity. Holding type-$H$ costs fixed, a higher $K_L$ makes unprofitable applicants less likely to complete, which weakly widens the recourse region at both ends. The movements are strict when costs increase strictly at every feasible requirement. Thus the recourse region can expand either because completion is more productive or because type-$L$ applicants are less likely to complete; the size of the region alone does not identify which channel drives its use.

A higher upper bound on the idiosyncratic cost shock has a different effect. Since $\varepsilon$ is uniform on $[0,\bar\varepsilon]$, increasing $\bar\varepsilon$ raises the mean shock as well as its dispersion; it is not a mean-preserving increase in uncertainty. With interior completion probabilities, the increase scales down the probability that either type completes any requirement. This common scaling does not change where the recourse payoff crosses zero, so $\underline p$ is unchanged. It does, however, make recourse less attractive relative to the positive payoff from outright acceptance at the upper cutoff, which lowers $\bar p$.

Finally, a larger loss $a_L$ makes low-posterior offers less attractive and therefore raises the lower cutoff. At the same time, recourse insulates the principal from some type-$L$ losses because only completers are accepted, so it becomes more attractive relative to unconditional acceptance and the upper cutoff also rises. A larger gain $a_H$ reverses both comparisons: recourse becomes profitable sooner, but outright acceptance captures the gain from every type-$H$ applicant rather than only those who complete, shifting both cutoffs to the left.


\section{The incidence of recourse}
\label{sec:incidence}

Allowing the principal to offer recourse changes the principal's decision on both sides of the no-recourse acceptance cutoff $p^A$. Some applicants who would otherwise be rejected receive a recourse offer, while some who would otherwise be accepted outright must complete a requirement to gain acceptance. How does allowing the principal to offer recourse affect applicant welfare?

Consider the no-recourse benchmark, in which the principal chooses only between acceptance and rejection. The tie-broken rule accepts when $p\ge p^A$ and rejects otherwise, so the applicant's expected payoff is
\[
W^0(p)=
\begin{cases}
0, & p<p^A,\\
v, & p\ge p^A.
\end{cases}
\]

If recourse with requirement $r$ is offered, an applicant of type $\theta$ obtains expected payoff
\[
B_\theta(r)\equiv\mathbb{E}_\varepsilon\big[\max\{v-K_\theta(r)-\varepsilon,\,0\}\big],
\]
that is, his expected surplus from being offered recourse, given that he walks away and gets nothing whenever his cost draw is too high to justify completing it. Conditional on posterior $p$, applicant welfare from the offer is $B(p,r)\equiv p B_H(r)+(1-p)B_L(r)$. Let $W^1(p)$ denote applicant welfare under the tie-broken optimal rule when recourse is available.
Under the threshold rule, the posterior scale divides into four intervals, on which recourse either leaves the benchmark outcome unchanged or moves it in a determinate direction.

\begin{proposition}[Redistribution across the acceptance margin]
\label{prop:redistribution}
Suppose Assumptions~\ref{ass:limited_recourse} and~\ref{ass:endpoint} hold, so that the optimal rule is the threshold rule of Proposition~\ref{prop:threshold} with cutoffs $\underline p<p^A<\bar p$. Then, relative to the no-recourse benchmark,
\[
W^1(p)=W^0(p)
\quad\text{for }p\in[0,\underline p)\cup[\bar p,1],
\]
\[
W^1(p)>W^0(p)=0
\quad\text{for }p\in[\underline p,p^A),
\]
\[
W^1(p)<W^0(p)=v
\quad\text{for }p\in[p^A,\bar p).
\]
\end{proposition}

Introducing recourse therefore need not benefit every applicant whose outcome changes. An evaluation confined to applicants who would otherwise be rejected captures the gains below $p^A$ but misses the losses above $p^A$, where recourse replaces outright acceptance. Applicant welfare must therefore be compared with the no-recourse outcome on both sides of the old acceptance cutoff.

Proposition~\ref{prop:comparative_statics} describes the extensive-margin effect of productivity: a higher $\delta$ expands the recourse region, giving some rejected applicants a conditional offer while requiring some previously accepted applicants to complete recourse. We now hold the posterior fixed within the recourse region and study the intensive margin: how productivity changes the requirement attached to an existing recourse offer. We consider local changes in $\delta$ that preserve the maintained assumptions, and hold fixed the applicant's value of acceptance and the cost of completing any given requirement.

Increasing the requirement has two screening effects. It discourages profitable applicants from completing recourse, which hurts the principal, but also discourages unprofitable applicants, which benefits the principal. Greater productivity changes how the principal trades off these screening effects against the value generated by a completed requirement.

\begin{proposition}[Productivity and the recourse requirement]
\label{prop:productivity_welfare}
Suppose Assumptions~\ref{ass:limited_recourse} and~\ref{ass:endpoint} hold. Fix a posterior $p\in(\underline p,\bar p)$ at which the optimal requirement is unique and lies in $(\underline r,\bar r)$. Suppose the principal's payoff has a strictly negative second derivative with respect to the requirement at this optimum, $U^O_{rr}(p,r^*(p))<0$. Holding $p$ fixed, a marginal increase in productivity $\delta$ raises the optimal requirement if
\[
p a_H K_H'(r^*(p))
>
(1-p)a_L K_L'(r^*(p)),
\]
and lowers it if the inequality is reversed. If the two sides are equal, the derivative of the optimal requirement with respect to productivity is zero: $\partial r^*(p)/\partial\delta=0$.
\end{proposition}

The left-hand side of the inequality is the expected marginal loss from discouraging profitable applicants: they occur with probability $p$, are worth $a_H$ to the principal, and become less likely to complete as their cost $K_H$ rises. The right-hand side is the expected marginal benefit from discouraging unprofitable applicants. When the loss from discouraging profitable applicants dominates, the productive return to a higher requirement supports the principal's choice of stringency. Raising $\delta$ strengthens this return, so the principal raises the requirement. When the screening benefit from discouraging unprofitable applicants dominates, a higher requirement reduces the expected value generated through completion. Raising $\delta$ makes that loss more costly, so the principal lowers the requirement.

Higher productivity $\delta$ does not directly benefit applicants: it leaves their reward from acceptance and the cost of a given requirement unchanged. What matters to applicants is whether the principal raises or lowers the requirement. A higher requirement reduces the surplus of those who complete recourse and causes some to withdraw. A lower requirement makes applicants better off.

This adds an intensive-margin source of applicant harm to the redistribution in Proposition~\ref{prop:redistribution}. Higher productivity can benefit applicants newly offered recourse while harming both those who lose outright acceptance and those whose existing recourse offer becomes more stringent. Thus an improvement that makes recourse more valuable to the principal need not improve the terms on which applicants gain acceptance.

\section{A linear-cost example}
\label{sec:linear_example}

To see how the threshold rule operates, consider linear recourse costs:
\[
K_\theta(r)=c_\theta r,
\qquad
0<c_H<c_L,
\]
so the same requirement is less costly for a profitable applicant than for an unprofitable applicant. In a lending application, $r$ could represent a required reduction in debt or increase in income. For the illustration, normalize $a_H=c_H=1$ and set $a_L=1.5$, $v=5$, $\delta=0.15$, $c_L=1.2$, $\bar\varepsilon=10$, $\underline r=0.5$, and $\bar r=2.8$. Appendix~\ref{app:linear_example} derives the solution. Readers can vary the parameters in an interactive version of the example, available at \url{https://matthewolckers.github.io/paper-recourse/}.

Without recourse, the principal accepts when $p\ge p^A=0.6$. With recourse, this single cutoff becomes a recourse region: the principal rejects below $\underline p=0.362$, offers recourse on $[0.362,0.661)$, and accepts outright at and above $\bar p=0.661$. Recourse therefore changes the decision only for applicants near the original accept--reject margin. Applicants with clearly unfavorable posteriors remain rejected, while those with clearly favorable posteriors remain accepted without conditions.

\begin{figure}[p]
    \centering
    \includegraphics[height=0.84\textheight,width=\textwidth,keepaspectratio]{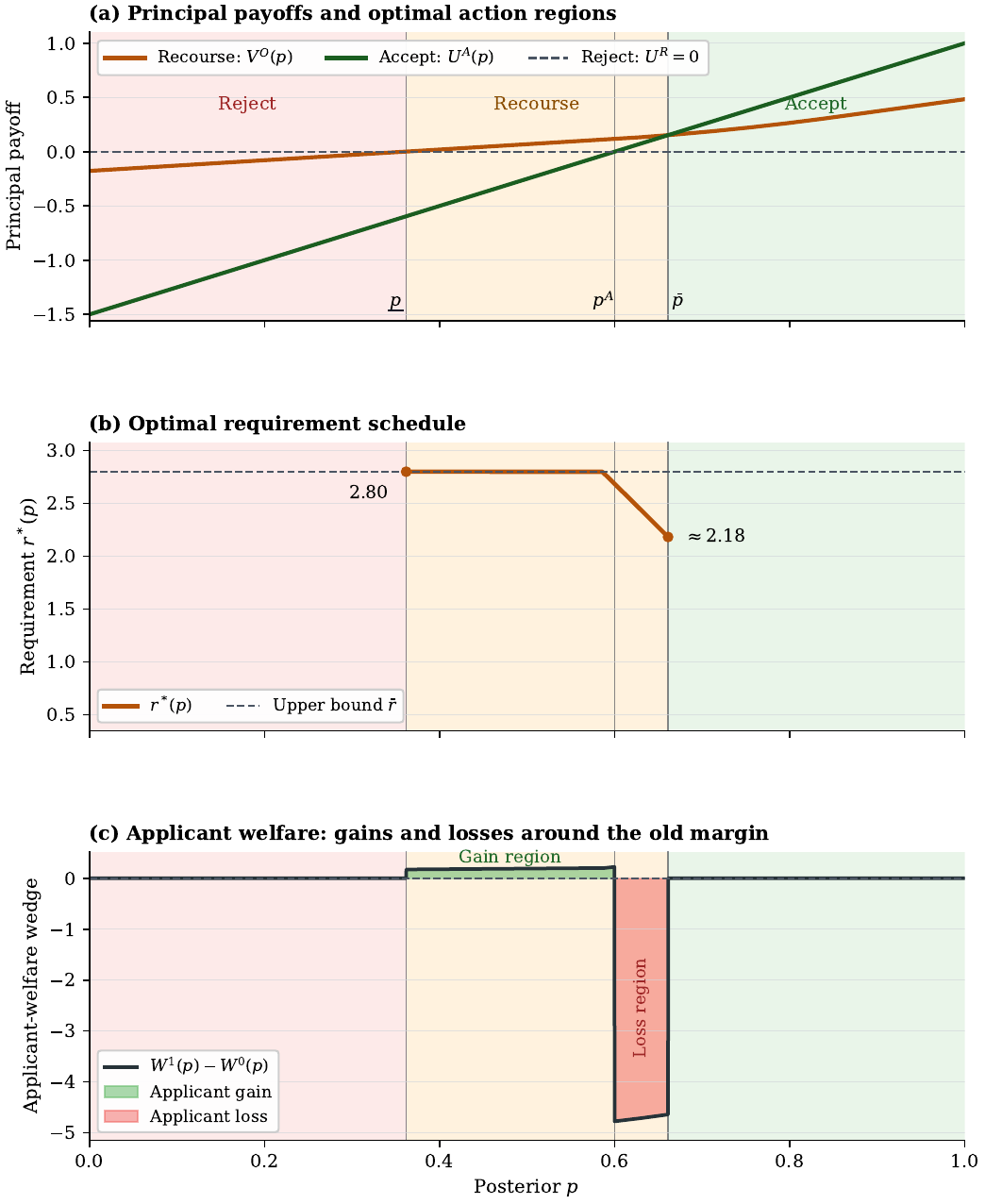}
    \caption{The linear-cost example. Panel~(a) compares the principal's payoff from optimized recourse, acceptance, and rejection. Panel~(b) shows the optimal requirement within the recourse region. Panel~(c) shows the change in applicant welfare relative to the no-recourse benchmark. In every panel, the background shading marks the optimal reject, recourse, and accept regions, and the vertical guides mark $\underline p$, $p^A$, and $\bar p$.}
    \label{fig:linear_example}
\end{figure}

Panel~(a) of Figure~\ref{fig:linear_example} explains why these three regions arise. The dashed horizontal line gives the payoff from rejection, the green line the payoff from immediate acceptance, and the orange curve the payoff from recourse when the requirement is chosen optimally. Immediate acceptance first becomes profitable at $p^A$, where the green line crosses zero. Recourse becomes profitable sooner: the orange curve crosses zero at $\underline p$. It remains the principal's preferred action until $\bar p$, where the green line overtakes it. The shaded regions are therefore determined by two simple comparisons: recourse versus rejection at the lower cutoff, and recourse versus immediate acceptance at the upper cutoff. Within the middle region, the vertical distance between the orange curve and the no-recourse payoff $\max\{U^A(p),0\}$ measures the principal's gain from offering recourse. Because $p^A$ lies inside this region, recourse is profitable both when it gives an otherwise rejected applicant a path to acceptance and when it adds a condition to an applicant who would otherwise be accepted immediately.

Panel~(b) shows how demanding the recourse offer is. At the lower cutoff, the principal sets the requirement at its maximum feasible level, $\bar r=2.8$. The initial flat segment indicates that this upper bound continues to bind: absent the bound, the principal would set an even tougher requirement for these relatively unfavorable applicants. At about $p=0.586$, the bound ceases to bind and the schedule begins to fall, reaching approximately $2.18$ at the upper cutoff. Thus applicants with less favorable posteriors face the tougher hurdle. Stringency is most valuable there because it discourages unprofitable applicants, who have higher completion costs, from taking up the offer. As the posterior improves, this screening benefit becomes less important and the principal relaxes the requirement.

The downward slope in panel~(b) describes how the requirement varies with the posterior while productivity is held fixed. Proposition~\ref{prop:productivity_welfare} instead varies productivity while holding the posterior fixed. With linear costs and an interior optimum, the proposition says that higher productivity raises the requirement exactly when
\[
p>
\frac{a_Lc_L}{a_Hc_H+a_Lc_L}.
\]
In this calibration, the effect of productivity on the requirement changes sign at approximately $p=0.643$. Increasing $\delta$ lowers the requirement at $p=0.60$ but raises it at $p=0.65$, with applicant welfare increasing in the first case and decreasing in the second. Greater productivity can therefore lead to a stricter requirement even though a more favorable posterior leads to a less demanding one.

Panel~(c) shows what the change in policy means for applicants. The plotted wedge, $W^1(p)-W^0(p)$, compares applicant welfare with and without recourse. It is zero outside the recourse region because the principal takes the same action under both regimes: rejection below $\underline p$ and immediate acceptance above $\bar p$. Between $\underline p$ and $p^A$, the wedge is positive. These applicants would otherwise be rejected, so even a costly, conditional route to acceptance makes them better off. At $p^A$, the no-recourse rule switches discontinuously from rejection to immediate acceptance, raising benchmark welfare from zero to $v$. The recourse offer itself does not change discontinuously there, which explains the sharp drop in the welfare wedge. Between $p^A$ and $\bar p$, the wedge is negative because applicants who would have been accepted immediately must instead complete a costly requirement. It returns to zero at $\bar p$, where immediate acceptance resumes. The larger magnitude of the losses in this calibration reflects this difference in benchmarks: recourse creates a chance of acceptance below $p^A$, but replaces certain, costless acceptance above it.

The example illustrates why the principal's incentive to offer recourse does not imply that every affected applicant benefits. The same policy that expands opportunity just below the original acceptance cutoff imposes a new hurdle just above it.

\section{Conclusion}

Algorithmic recourse promises to turn rejection into an opportunity for acceptance. By specifying concrete steps an applicant can take, recourse offers a path forward after an unfavorable decision. Yet allowing decision-makers to offer recourse also changes the decision-maker's choice between outright acceptance, rejection, and a conditional offer.

Our main result is that recourse can harm applicants. Making recourse available can induce the principal to impose a costly requirement on applicants who would otherwise be accepted outright. Under the threshold rule, applicants just below the original acceptance cutoff gain a chance at acceptance, while those just above it must complete recourse to obtain an outcome previously available without that requirement. Recourse therefore changes both who can gain acceptance and the terms on which acceptance is offered.

Making recourse more productive need not resolve this conflict. Higher productivity expands the recourse region, benefiting some previously rejected applicants but imposing requirements on some previously accepted applicants. It can also induce the principal to make existing recourse offers more stringent, reducing the welfare of applicants already offered recourse. Improvements that make recourse more valuable to the principal can thus make applicants worse off.

Our model opens several avenues for future work. We have considered a single principal setting a decision rule, but often the decision-maker is in competition for applicants. In a competitive environment, a rival's offer changes the applicant's outside option: the relevant comparison is no longer recourse versus zero, but recourse versus applying elsewhere. Whether competition disciplines the region $[p^A,\bar p)$ in which recourse harms applicants, or strengthens incentives to offer recourse below $p^A$, remains an open question.

These strategic and distributional findings raise a policy question. Because a principal offers recourse only when it serves her own objective, universal accessibility is not guaranteed. When recourse can neither compensate for type-$L$ losses nor screen out type-$L$ applicants completely, sufficiently unfavorable applicants remain rejected. This gap may fuel pressure to make algorithmic recourse mandatory in high-stakes domains. Studying mandatory recourse is a natural next step, and could be formalized as a game between a regulator, firms, and a population of applicants. A regulator might set a floor on recourse accessibility or a ceiling on requirement stringency; firms would respond strategically, perhaps by adjusting baseline classifiers, tightening initial screening criteria, or shifting costs elsewhere. The welfare results here suggest that mandates should be evaluated not only by whether they increase access to recourse, but also by which applicants gain or lose relative to the old acceptance margin.

\bibliographystyle{jpe}
\bibliography{references.bib}

\appendix

\section{Proofs}

\begin{proof}[Proof of Proposition~\ref{prop:compensation_screening}]
Fix a posterior $p\in[0,1]$. The principal's problem is pointwise in $p$. Acceptance yields
\[
U^A(p)
=
p a_H-(1-p)a_L
=
p(a_H+a_L)-a_L.
\]
Thus acceptance weakly dominates rejection if and only if
\[
p\ge p^A\equiv \frac{a_L}{a_H+a_L}.
\]
At $p=0$, acceptance is strictly worse than rejection:
\[
U^A(0)=-a_L<0.
\]

Now consider recourse with requirement $r\in\mathcal R$. Its expected payoff is
\[
U^O(p,r)
=
p q_H(r)(a_H+\delta r)
+
(1-p)q_L(r)(\delta r-a_L).
\]
Since $q_H(r)\ge 0$,
\[
q_H(r)(a_H+\delta r)\ge 0.
\]
Therefore, as a function of $p$, $U^O(p,r)$ is affine and has a nonnegative value at $p=1$. A recourse requirement $r$ weakly dominates rejection at every posterior if and only if
\begin{equation}
U^O(0,r)
=
q_L(r)(\delta r-a_L)
\ge 0.
\label{eq:proof_key_condition}
\end{equation}
Indeed, if \eqref{eq:proof_key_condition} holds, both endpoint values $U^O(0,r)$ and $U^O(1,r)$ are nonnegative, so the affine function $U^O(\cdot,r)$ is nonnegative on all of $[0,1]$. Conversely, if recourse weakly dominates rejection at every posterior, it must do so at $p=0$, giving \eqref{eq:proof_key_condition}.

Condition \eqref{eq:proof_key_condition} has exactly two interpretations. If $q_L(r)>0$, then $K_L(r)<v$, and \eqref{eq:proof_key_condition} is equivalent to
\[
\delta r-a_L\ge 0,
\]
or $r\ge \frac{a_L}{\delta}$. This is condition \textup{(C)}. If instead $q_L(r)=0$, then $K_L(r)\ge v$. The requirement screens out type $L$, and $U^O(p,r)=p\,q_H(r)(a_H+\delta r)\ge 0=U^R$ for all $p$; when payoffs are equal, the tie-breaking rule selects recourse over rejection. This is condition \textup{(S)}.

We can now prove sufficiency. If \textup{(C)} holds, choose a requirement $r$ satisfying \eqref{eq:cond_compensation}. Then $q_L(r)>0$ and $\delta r-a_L\ge 0$, so
\[
U^O(0,r)=q_L(r)(\delta r-a_L)\ge 0.
\]
Hence, by the preceding argument,
\[
U^O(p,r)\ge 0=U^R
\qquad\text{for every }p\in[0,1].
\]
If \textup{(S)} holds, choose $r$ with $K_L(r)\ge v$. Then $q_L(r)=0$. Hence
\[
U^O(p,r)
=
p q_H(r)(a_H+\delta r)
\ge 0
\qquad\text{for every }p\in[0,1].
\]
In both cases, recourse weakly dominates rejection at every posterior. Since the tie-breaking rule selects recourse over rejection whenever the two yield the same payoff, rejection is never selected.

For necessity, suppose rejection is globally redundant. At posterior $p=0$, acceptance yields $-a_L<0$, so acceptance cannot displace rejection. Hence some recourse requirement $r$ must satisfy
\[
U^O(0,r)=q_L(r)(\delta r-a_L)\ge 0.
\]
If $q_L(r)>0$, then $K_L(r)<v$ and necessarily $\delta r\ge a_L$, so \textup{(C)} holds. If $q_L(r)=0$, then $K_L(r)\ge v$, so \textup{(S)} holds. Thus global redundancy of rejection implies \textup{(C)} or \textup{(S)}.

The two conditions are therefore necessary and sufficient.
\end{proof}


\begin{proof}[Proof of Proposition~\ref{prop:threshold}]
Because $\mathcal R$ is compact and $U^O$ is continuous in $r$, the maximum in the definition of $V^O$ is attained for every $p$. The functions $U^O(\cdot,r)$ have slopes bounded uniformly in $r$, so their maximum $V^O$ is continuous.

We first establish the shape of the value function. For each fixed $r\in\mathcal R$, the payoff $U^O(\cdot,r)$ in~\eqref{eq:recourse_payoff} is affine in $p$. Hence, for any $p_1,p_2\in[0,1]$ and any $\lambda\in[0,1]$,
\[
\begin{aligned}
V^O(\lambda p_1+(1-\lambda)p_2)
&=\max_{r\in\mathcal R}
\left\{\lambda U^O(p_1,r)+(1-\lambda)U^O(p_2,r)\right\} \\
&\le
\lambda V^O(p_1)+(1-\lambda)V^O(p_2).
\end{aligned}
\]
Thus $V^O$ is convex. Moreover, the cost ordering~\eqref{ass:cost_ordering} and Assumption~\ref{ass:limited_recourse} imply $K_H(\bar r)\le K_L(\bar r)<v$. Therefore
\[
\eta_H\equiv \min\left\{\frac{v-K_H(\bar r)}{\bar\varepsilon},1\right\}>0
\]
is a uniform lower bound on $q_H(r)$. The slope of $U^O(\cdot,r)$ is
\[
q_H(r)(a_H+\delta r)+q_L(r)(a_L-\delta r).
\]
Because $\delta r\le\delta\bar r<a_L$, this slope is at least $\eta_Ha_H>0$. Thus, if $p_2>p_1$,
\[
U^O(p_2,r)\ge U^O(p_1,r)+\eta_Ha_H(p_2-p_1)
\qquad\text{for every }r\in\mathcal R.
\]
Taking maxima over $r$ shows that $V^O$ is strictly increasing.

We next establish that recourse is strictly valuable at the ordinary acceptance cutoff. At $p^A=a_L/(a_H+a_L)$,
\[
p^Aa_H=(1-p^A)a_L\equiv m.
\]
Substituting into~\eqref{eq:recourse_payoff} gives
\[
U^O(p^A,r)
=m\big[q_H(r)-q_L(r)\big]
+\delta r\big[p^Aq_H(r)+(1-p^A)q_L(r)\big].
\]
The cost ordering and the independence of $\varepsilon$ from type imply $q_H(r)\ge q_L(r)$, so the first term is nonnegative. In addition, $K_H(r)\le K_L(\bar r)<v$, so $q_H(r)>0$. Since $r\ge\underline r>0$, the second term is strictly positive. Hence $U^O(p^A,r)>0$ for every feasible $r$, and therefore $V^O(p^A)>0$.

First compare recourse with rejection. Under Assumption~\ref{ass:limited_recourse}, every feasible $r$ satisfies $K_L(r)\le K_L(\bar r)<v$ and $\delta r\le \delta\bar r<a_L$. Define
\[
\eta_L\equiv \min\left\{\frac{v-K_L(\bar r)}{\bar\varepsilon},1\right\}>0
\qquad\text{and}\qquad
\gamma\equiv a_L-\delta\bar r>0 .
\]
Then $q_L(r)\ge \eta_L$ and $a_L-\delta r\ge\gamma$ for every $r\in\mathcal R$, so
\[
U^O(0,r)=q_L(r)(\delta r-a_L)\le -\eta_L\gamma<0
\qquad\text{for every }r\in\mathcal R.
\]
Thus $V^O(0)<0$. Since $V^O(p^A)>0$ and $V^O$ is continuous and strictly increasing, there is a unique $\underline p\in(0,p^A)$ such that
\[
V^O(\underline p)=0,\qquad
V^O(p)<0\ \text{for }p<\underline p,\qquad
V^O(p)\ge 0\ \text{for }p\ge\underline p.
\]

Next compare recourse with acceptance. Let
\[
\Delta(p)\equiv V^O(p)-U^A(p).
\]
Since $V^O$ is convex and $U^A$ is affine, $\Delta$ is convex and continuous. At the ordinary acceptance cutoff, $U^A(p^A)=0$, so
\[
\Delta(p^A)=V^O(p^A)>0.
\]
At $p=1$, Assumption~\ref{ass:endpoint} gives
\[
\Delta(1)
=V^O(1)-a_H
=\max_{r\in\mathcal R}\{q_H(r)(a_H+\delta r)\}-a_H
\le 0.
\]
The sublevel set $\{p\in[0,1]:\Delta(p)\le 0\}$ is therefore a nonempty closed interval containing $1$ and excluding $p^A$. Hence there is a cutoff $\bar p\in(p^A,1]$ such that
\[
\Delta(p)>0\ \text{for }p<\bar p,
\qquad
\Delta(p)\le 0\ \text{for }p\ge\bar p.
\]

The acceptance region has positive width precisely when the endpoint comparison is strict. If $V^O(1)<a_H$, continuity gives $\bar p<1$. Conversely, if $V^O(1)=a_H$, compactness implies that some $r_0\in\mathcal R$ satisfies $q_H(r_0)(a_H+\delta r_0)=a_H$. For every $p<1$,
\[
U^O(p,r_0)-U^A(p)
=
(1-p)\big[a_L(1-q_L(r_0))+\delta r_0q_L(r_0)\big]
>0.
\]
Thus acceptance is not optimal at any $p<1$, so $\bar p=1$.

We can now compare all three actions. If $p<\underline p$, then $p<p^A$, so $U^A(p)<0$, and also $V^O(p)<0$; rejection is strictly optimal. If $\underline p\le p<\bar p$, then $V^O(p)\ge0=U^R$ and $\Delta(p)>0$, so recourse weakly beats rejection and strictly beats acceptance. At $p=\underline p$, any tie is between recourse and rejection, and the tie-breaking rule selects recourse. If $p\ge\bar p$, then $p>p^A$, so $U^A(p)>0=U^R$, and $\Delta(p)\le0$, so acceptance weakly beats recourse and strictly beats rejection. Any tie between acceptance and recourse is resolved in favor of acceptance.

Finally, continuity of $U^O$ and compactness of $\mathcal R$ imply, by the measurable maximum theorem, that a measurable maximizing selection $r^*(p)$ exists on $[\underline p,\bar p)$. This selection defines a measurable optimal rule $\varphi^*$. Since $p(s)$ is measurable, the action following signal $s$ is given by the measurable rule $\varphi^*(p(s))$.
\end{proof}


\begin{proof}[Proof of Proposition~\ref{prop:redistribution}]
In the no-recourse benchmark, the tie-broken rule rejects for $p<p^A$ and accepts for $p\ge p^A$. Under Proposition~\ref{prop:threshold}, the rule with recourse rejects for $p<\underline p$, offers recourse for $\underline p\le p<\bar p$, and accepts for $p\ge\bar p$.

On $[0,\underline p)$ both rules reject, so $W^1(p)=W^0(p)=0$. On $[\bar p,1]$ both rules accept outright, so $W^1(p)=W^0(p)=v$.

Now take $p\in[\underline p,p^A)$. The benchmark rejects, so $W^0(p)=0$. The rule with recourse offers the requirement $r^*(p)\in\mathcal R$. For any feasible requirement $r$, Assumption~\ref{ass:limited_recourse} and the cost ordering~\eqref{ass:cost_ordering} imply
\[
K_H(r)\le K_L(r)\le K_L(\bar r)<v.
\]
Thus $v-K_\theta(r)>0$ for both types. Since $\varepsilon$ has a continuous uniform distribution on $[0,\bar\varepsilon]$, the event
\[
\varepsilon<\min\{v-K_\theta(r),\bar\varepsilon\}
\]
has positive probability, and on that event $v-K_\theta(r)-\varepsilon>0$. Therefore $B_\theta(r)>0$ for both $\theta=H,L$. It follows that
\[
W^1(p)=pB_H(r^*(p))+(1-p)B_L(r^*(p))>0=W^0(p).
\]

Finally take $p\in[p^A,\bar p)$. The benchmark accepts outright, so $W^0(p)=v$, while the rule with recourse offers $r^*(p)\in\mathcal R$. By the maintained restrictions on the cost function, $K_\theta(r^*(p))\ge0$ for each type, and $\varepsilon>0$ almost surely. Hence
\[
\max\{v-K_\theta(r^*(p))-\varepsilon,\,0\}\le \max\{v-\varepsilon,\,0\}<v
\qquad\text{almost surely}.
\]
Taking expectations gives $B_\theta(r^*(p))<v$ for both types. Therefore
\[
W^1(p)=pB_H(r^*(p))+(1-p)B_L(r^*(p))<v=W^0(p).
\]
\end{proof}


\begin{proof}[Proof of Proposition~\ref{prop:comparative_statics}]
The maintained assumptions imply
\[
0<q_L(r)\le q_H(r)<1
\qquad\text{for every }r\in\mathcal R.
\]
We compare payoffs for each fixed requirement and then maximize over the same feasible set. Weak pointwise comparisons are preserved by maximization. Strict pointwise comparisons are also preserved here because the payoff functions are continuous and their maxima on the compact set $\mathcal R$ are attained. The sign patterns in Proposition~\ref{prop:threshold} then translate these comparisons into movements of the cutoffs.

For productivity, at every posterior and feasible requirement,
\[
\frac{\partial U^O(p,r)}{\partial\delta}
=
r\big[pq_H(r)+(1-p)q_L(r)\big]>0.
\]
A higher $\delta$ therefore strictly raises $V^O(p)$, while leaving $U^A(p)$ unchanged. At the old lower cutoff, the new recourse value is strictly positive, so $\underline p$ falls. At the old upper cutoff, the new recourse value strictly exceeds acceptance, so $\bar p$ rises.

Next, suppose type-$L$ costs change from $K_L$ to $\widetilde K_L$, where $\widetilde K_L(r)\ge K_L(r)$ for every feasible $r$, and let $\widetilde U^O$ denote the corresponding recourse payoff. Interior completion probabilities imply
\[
\widetilde U^O(p,r)-U^O(p,r)
=
\frac{(1-p)(a_L-\delta r)}{\bar\varepsilon}
\big[\widetilde K_L(r)-K_L(r)\big]
\ge0.
\]
The inequality uses $\delta r<a_L$. Maximizing therefore weakly raises the recourse value while leaving acceptance unchanged, so the lower cutoff weakly falls and the upper cutoff weakly rises. If $\widetilde K_L(r)>K_L(r)$ for every feasible $r$, the payoff comparison is strict at every $p<1$, including both original cutoffs. Both cutoff movements are then strict.

For the cost-shock upper bound, interior completion probabilities give
\[
\bar\varepsilon V^O(p)
=
\max_{r\in\mathcal R}
\left\{
p[v-K_H(r)](a_H+\delta r)
+
(1-p)[v-K_L(r)](\delta r-a_L)
\right\}.
\]
The expression on the right does not depend on $\bar\varepsilon$. Increasing $\bar\varepsilon$ therefore multiplies the recourse value by a common positive factor smaller than one. This leaves its zero, and hence $\underline p$, unchanged. At the old upper cutoff, $V^O(\bar p)=U^A(\bar p)>0$, so the scaling strictly lowers recourse relative to acceptance. The upper cutoff therefore falls.

Finally, for any fixed requirement and $p\in(0,1)$,
\[
\begin{aligned}
\frac{\partial U^O(p,r)}{\partial a_L}
&=-(1-p)q_L(r)<0,\\
\frac{\partial[U^O(p,r)-U^A(p)]}{\partial a_L}
&=(1-p)[1-q_L(r)]>0.
\end{aligned}
\]
A higher $a_L$ therefore strictly lowers the optimized recourse payoff relative to rejection and strictly raises it relative to acceptance. Both cutoffs rise. Similarly,
\[
\begin{aligned}
\frac{\partial U^O(p,r)}{\partial a_H}
&=p q_H(r)>0,\\
\frac{\partial[U^O(p,r)-U^A(p)]}{\partial a_H}
&=-p[1-q_H(r)]<0.
\end{aligned}
\]
A higher $a_H$ strictly raises recourse relative to rejection and strictly lowers it relative to acceptance. Both cutoffs fall.
\end{proof}



\begin{proof}[Proof of Proposition~\ref{prop:productivity_welfare}]
Fix $p$ and suppress the argument $p$ of $r^*$. The principal's payoff from recourse can be written as
\[
U^O(p,r)
=
p a_H q_H(r)
-(1-p)a_L q_L(r)
+\delta r\big[p q_H(r)+(1-p)q_L(r)\big].
\]
The first two terms capture selection into completion, while the last term is the value generated by recourse.

The maintained assumptions imply interior completion probabilities throughout the feasible set, so
\[
q_\theta'(r)=-\frac{K_\theta'(r)}{\bar\varepsilon}.
\]
Using this expression in the first-order condition $U^O_r(p,r^*)=0$ gives
\[
\delta U^O_{r\delta}(p,r^*)
=
\frac{
    p a_H K_H'(r^*)
    -(1-p)a_L K_L'(r^*)
}{\bar\varepsilon}.
\]
Uniqueness, interiority, and $U^O_{rr}(p,r^*)<0$ allow local differentiation of the first-order condition. Because $p$ lies strictly inside the recourse region, a sufficiently small change in $\delta$ leaves recourse strictly optimal. Implicit differentiation gives
\[
\frac{\partial r^*(p)}{\partial\delta}
=
\frac{
    p a_H K_H'(r^*(p))
    -(1-p)a_L K_L'(r^*(p))
}{
    -\delta\bar\varepsilon U^O_{rr}(p,r^*(p))
}.
\]
The denominator is positive, so the numerator determines whether the optimal requirement rises or falls.

Finally, applicant welfare is strictly decreasing in the requirement. In particular,
\[
B_\theta'(r)
=
-q_\theta(r)K_\theta'(r)
<0.
\]
At the fixed posterior, $\delta$ affects applicants only through the requirement chosen by the principal. Applicant welfare therefore moves in the opposite direction to $r^*$, separately for each type and for their posterior-weighted average.
\end{proof}

\section{Additional results}

\subsection{Details for the linear-cost example}
\label{app:linear_example}

This appendix records the calculations used to construct Figure~\ref{fig:linear_example}. Under the linear specification $K_\theta(r)=c_\theta r$, suppose completion probabilities are interior on the feasible set:
\[
0<v-c_L\bar r
\qquad\text{and}\qquad
v-c_H\underline r<\bar\varepsilon .
\]
Then $q_\theta(r)=(v-c_\theta r)/\bar\varepsilon$, and the principal's payoff from recourse is
\begin{equation}
U^O(p,r)
=
\frac{p}{\bar\varepsilon}(v-c_H r)(a_H+\delta r)
+
\frac{1-p}{\bar\varepsilon}(v-c_L r)(\delta r-a_L).
\label{eq:linear_UO}
\end{equation}
Define
\[
C(p)\equiv p c_H+(1-p)c_L
\qquad\text{and}\qquad
A(p)\equiv \delta v+(1-p)c_La_L-pc_Ha_H.
\]
Equation~\eqref{eq:linear_UO} can then be written as
\[
\bar\varepsilon U^O(p,r)
=vU^A(p)+A(p)r-\delta C(p)r^2 .
\]
The objective is strictly concave and quadratic in $r$. Its unconstrained maximizer is
\begin{equation}
\hat r(p)
=
\frac{\delta v+(1-p)c_La_L-pc_Ha_H}
{2\delta\big[p c_H+(1-p)c_L\big]},
\label{eq:linear_rhat}
\end{equation}
so the optimal feasible requirement is the projection
\[
r^*(p)=\min\{\bar r,\max\{\underline r,\hat r(p)\}\}.
\]
Substitution gives the optimized recourse value
\begin{equation}
V^O(p)
=
\frac{1}{\bar\varepsilon}
\left[
vU^A(p)+A(p)r^*(p)-\delta C(p)\big(r^*(p)\big)^2
\right].
\label{eq:linear_value_projected}
\end{equation}
When neither feasibility bound binds, this reduces to
\[
V^O(p)
=
\frac{1}{\bar\varepsilon}
\left[
vU^A(p)+\frac{A(p)^2}{4\delta C(p)}
\right].
\]
The lower cutoff solves $V^O(\underline p)=0$, the no-recourse cutoff is $p^A=a_L/(a_H+a_L)$, and the upper cutoff solves $V^O(\bar p)=U^A(\bar p)$.

For the calibration in Section~\ref{sec:linear_example}, the slack lower bound is $\underline r=0.5$. Substituting the calibrated primitives yields
\[
\underline p=0.362,
\qquad
p^A=0.6,
\qquad
\bar p=0.661,
\]
with $r^*(\underline p)=2.8$ and $r^*(\bar p)=2.182426$. Finally, applicant surplus from a recourse offer is
\[
B_\theta(r)
=
\mathbb E_\varepsilon\big[\max\{v-c_\theta r-\varepsilon,0\}\big]
=
\frac{(v-c_\theta r)^2}{2\bar\varepsilon},
\]
which generates the pointwise welfare wedge in panel~(c).

\subsection{Structure without the endpoint condition}

Assumption~\ref{ass:endpoint} is not needed for the lower reject--recourse cutoff. It determines whether outright acceptance is selected anywhere on the full posterior interval. The acceptance region may consist only of $p=1$; a region of positive width requires strict acceptance dominance at the top. If the weak condition fails, the same requirement that beats acceptance when the applicant is known to be type $H$ also beats acceptance when the applicant is known to be type $L$, and therefore beats acceptance at every posterior by linearity.

\begin{proposition}[Structure without the endpoint condition]
\label{prop:structure}
Suppose Assumption~\ref{ass:limited_recourse} holds. Then there is a cutoff $\underline p\in(0,p^A)$ such that rejection is optimal exactly on $[0,\underline p)$ and recourse is optimal on $[\underline p,p^A]$. The acceptance region is nonempty if and only if Assumption~\ref{ass:endpoint} holds. If Assumption~\ref{ass:endpoint} fails, acceptance is never optimal and the tie-broken optimal rule reduces to two regions: reject below $\underline p$ and offer recourse at and above $\underline p$.
\end{proposition}

\begin{proof}[Proof of Proposition~\ref{prop:structure}]
The proof of Proposition~\ref{prop:threshold} did not use Assumption~\ref{ass:endpoint} to construct the lower cutoff. Under Assumption~\ref{ass:limited_recourse}, $V^O$ is continuous and strictly increasing, $V^O(0)<0$, and $V^O(p^A)>0$. Hence there is a unique $\underline p\in(0,p^A)$ such that rejection is strictly optimal for $p<\underline p$ and recourse weakly beats rejection for $p\ge\underline p$. Since $U^A(p)<0$ for $p<p^A$ and $V^O(p^A)>U^A(p^A)$, recourse is optimal throughout $[\underline p,p^A]$, with the boundary tie at $\underline p$ resolved in favor of recourse.

It remains to characterize acceptance. For a fixed requirement $r$, define the payoff difference between offering that requirement and accepting outright:
\[
D_r(p)\equiv U^O(p,r)-U^A(p).
\]
Using~\eqref{eq:accept_payoff} and~\eqref{eq:recourse_payoff},
\[
D_r(p)
=p\big[q_H(r)(a_H+\delta r)-a_H\big]
+(1-p)\big[a_L(1-q_L(r))+q_L(r)\delta r\big].
\]
The second bracket is strictly positive for every feasible $r$: if $q_L(r)=1$, it equals $\delta r>0$, while if $q_L(r)<1$, the term $a_L(1-q_L(r))$ is strictly positive and $q_L(r)\delta r$ is nonnegative.

If Assumption~\ref{ass:endpoint} fails, then there is some $r_0\in\mathcal R$ such that
\[
q_H(r_0)(a_H+\delta r_0)-a_H>0.
\]
For this same requirement, both endpoint values of the affine function $D_{r_0}$ are strictly positive, so $D_{r_0}(p)>0$ for every $p\in[0,1]$. Therefore
\[
V^O(p)-U^A(p)\ge D_{r_0}(p)>0
\qquad\text{for every }p\in[0,1],
\]
and acceptance is never optimal. Since rejection is optimal only below $\underline p$, recourse is optimal at every posterior $p\ge\underline p$.

Conversely, if Assumption~\ref{ass:endpoint} holds, Proposition~\ref{prop:threshold} applies and gives a nonempty acceptance region $[\bar p,1]$ with $\bar p\le1$. Thus the acceptance region is nonempty exactly under Assumption~\ref{ass:endpoint}.
\end{proof}

Proposition~\ref{prop:structure} characterizes when recourse displaces outright acceptance. If Assumption~\ref{ass:endpoint} fails, the productivity benefit makes a conditional offer preferable even when the applicant is known to be profitable. Under limited feasible recourse, the principal then offers recourse to every applicant whose posterior is at least $\underline p$. This does not eliminate the screening role of the requirement: whenever $q_H(r)>q_L(r)$, completion still selects a more profitable group of applicants. The result concerns the disappearance of unconditional acceptance, not the disappearance of screening.

Combining a condition that eliminates rejection in Proposition~\ref{prop:compensation_screening} with the failure of Assumption~\ref{ass:endpoint} allows recourse to displace both rejection and outright acceptance. The resulting policy maps every posterior into a conditional offer rather than reserving recourse for applicants near an acceptance margin. Every applicant faces a requirement, but some applicants may never find completion worthwhile. Universal provision of conditional offers therefore need not imply a positive probability of acceptance for every applicant.

This interpretation resembles score-based mortgage underwriting schedules in the United States, where an applicant's credit score maps into a loan decision or a set of terms, and improving the score may move the applicant to more favorable terms.\footnote{We thank Meredith Paker for drawing our attention to this connection.} The analogy concerns the form of the policy---a score-contingent schedule of conditional offers---rather than a claim that mortgage underwriting is literally a recourse mechanism. In particular, the present model does not by itself imply that the optimal requirement $r^*(p)$ becomes less demanding as the score rises; that monotonicity would require additional assumptions.

\subsection{Commitment and acceptance after completion}

The principal commits to a decision rule before observing the signal. Two aspects of this rule remain optimal at later stages: the prescribed initial action maximizes the principal's payoff at each posterior, and accepting an applicant who completes recourse is weakly optimal. These observations do not establish that the entire policy is self-enforcing, because rejecting an applicant who does not complete recourse need not be optimal.

\begin{proposition}[Pointwise optimality and acceptance after completion]
\label{prop:self_enforcing}
Suppose the threshold rule in Proposition~\ref{prop:threshold} applies. Holding fixed the continuation decisions specified by each action, the ex ante optimal rule also maximizes the principal's payoff after the signal is observed. Moreover, under the completion behavior induced by the committed rule, accepting a completer is weakly optimal relative to rejection at every posterior in the recourse region.
\end{proposition}

\begin{proof}[Proof of Proposition~\ref{prop:self_enforcing}]
The principal's objective is additively separable across signals, and the posterior $p(s)$ is a sufficient statistic for payoffs. Holding fixed the implementation of each action, the ex ante optimal rule therefore solves the same pointwise problem as the principal faces after observing the signal: choose the largest of $U^A(p)$, $0$, and $V^O(p)$.

Now fix a posterior $p\in[\underline p,\bar p)$ and a selected optimal requirement $r^*(p)$. For any feasible requirement, let
\[
Q(p,r)\equiv p q_H(r)+(1-p)q_L(r)
\]
denote the probability of completion. The maintained assumptions imply $0<q_L(r)\le q_H(r)<1$, and hence $0<Q(p,r)<1$. Conditional on completion, the principal's expected payoff from accepting the applicant is
\[
\frac{U^O(p,r^*(p))}{Q(p,r^*(p))}
=
\frac{V^O(p)}{Q(p,r^*(p))}
\ge 0,
\]
because recourse is selected only where $V^O(p)\ge0$. Rejecting a completer yields zero. Thus accepting a completer is weakly optimal.
\end{proof}

The continuation decision after noncompletion is different. If the principal could instead accept a non-completer, the expected payoff from doing so would be
\[
\frac{
p[1-q_H(r)]a_H-(1-p)[1-q_L(r)]a_L
}{
1-Q(p,r)
}
=
\frac{
U^A(p)-U^O(p,r)+\delta rQ(p,r)
}{
1-Q(p,r)
}.
\]
This payoff can be positive even when recourse is preferred to immediate acceptance: the principal may favor recourse for its productivity benefit rather than because non-completers are unprofitable.

Indeed, under the threshold rule, accepting a non-completer is strictly profitable at recourse posteriors sufficiently close to $\bar p$. To see this, substitute $r=r^*(p)$ and $U^O(p,r^*(p))=V^O(p)$ into the expression above. As $p\uparrow\bar p$, continuity implies $V^O(p)-U^A(p)\to0$. In contrast, $\delta r^*(p)Q(p,r^*(p))$ is bounded away from zero because $\underline r>0$ and completion probabilities have a uniform positive lower bound. The numerator is therefore positive sufficiently close to $\bar p$.

Commitment to reject non-completers consequently remains substantive in the present model. If withdrawal instead irrevocably terminates the application, accepting a non-completer is not an available deviation, but that is an additional restriction on continuation actions. The proposition establishes credibility of acceptance after completion, not credibility of the entire policy without commitment.

\subsection{Pure screening}
\label{sec:pure_screening}

This subsection relaxes the maintained restriction $\delta>0$ and considers the boundary case $\delta=0$, which isolates recourse as a pure test. It is not degenerate. Instead, it shows exactly what recourse can do when recommended actions do not improve the applicant's underlying value to the principal.

When $\delta=0$, no finite requirement can satisfy the primitive compensation condition $\delta r\ge a_L$. Rejection can be globally redundant only through screening: some feasible requirement must screen type $L$ applicants out completely. Thus purely performative recourse can displace rejection only if it is a sufficiently sharp screen.

At the ordinary accept--reject margin, the recourse payoff becomes
\[
U^O(p^A,r)
=m\big[q_H(r)-q_L(r)\big],
\qquad
m\equiv p^A a_H=(1-p^A)a_L .
\]
Thus pure-screening recourse has bite at the margin if and only if completion is informative, in the sense that $q_H(r)>q_L(r)$ for some feasible requirement. With the uniform shock and interior completion probabilities, this is equivalent to a strict cost ranking $K_H(r)<K_L(r)$ at that requirement. If $K_H=K_L$ on the feasible set, recourse neither improves applicants nor separates types, and it has no value at the margin.

Finally, Assumption~\ref{ass:endpoint} holds automatically when $\delta=0$, because it reduces to $q_H(r)a_H\le a_H$. Hence the case in which recourse displaces acceptance everywhere is a genuinely productive-recourse phenomenon: it requires $\delta$ large enough that the principal wants to extract the productivity bonus even from applicants known to be type $H$. By continuity, the pure-screening conclusions extend to small positive values of $\delta$ whenever the relevant payoff comparisons are strict.

\section{A screening formulation}
\label{app:screening}
This appendix states the principal's optimization problem together with the applicant's incentive and participation constraints. It then shows how these constraints yield the completion probabilities and the reduced-form objective used in the main text. The model is a screening problem with a restricted contract space: there are no monetary transfers, acceptance gives the applicant a fixed benefit $v$, and the principal can offer a single requirement whose completion guarantees acceptance. Screening operates through the applicant's decision to complete the requirement or withdraw.

\paragraph{Feasible mechanisms and information}
A mechanism is a measurable decision rule
\[
    \varphi:[0,1]\longrightarrow\mathcal A,
    \qquad
    \mathcal A
    =
    \{A,R\}\cup\bigl(\{O\}\times\mathcal R\bigr),
\]
where $A$ denotes unconditional acceptance, $R$ denotes rejection, and $(O,r)$ denotes a recourse offer with requirement $r\in\mathcal R=[\underline r,\bar r]$. Following signal $s$, the implemented action is $\varphi(p(s))$. The principal commits to $\varphi$ before observing the signal. Unconditional acceptance gives the applicant $v$ without any recourse cost. Under $(O,r)$, completion leads to acceptance and withdrawal leads to rejection. Thus the commitment includes both acceptance after completion and rejection after withdrawal.

The applicant privately observes his type $\theta\in\{H,L\}$. The principal observes the signal $s$ and forms the posterior $p(s)=\Pr(\theta=H\mid s)$. If recourse is offered, the applicant observes $r$ and then privately observes the cost shock $\varepsilon\sim\operatorname{Unif}[0,\bar\varepsilon]$, which is independent of $\theta$ and $s$. The principal cannot condition the offer directly on $\theta$ or $\varepsilon$. In particular, at a given signal the same requirement applies regardless of the applicant's privately observed type and subsequent cost realization.

\paragraph{The applicant's incentive and participation constraints}
Fix a recourse offer $r$. Let $b_\theta(\varepsilon;r)\in\{0,1\}$ denote the applicant's completion decision, with $b_\theta(\varepsilon;r)=1$ indicating completion and $b_\theta(\varepsilon;r)=0$ indicating withdrawal. His continuation payoff is
\[
    u_\theta(b;r,\varepsilon)
    =
    b\bigl[v-K_\theta(r)-\varepsilon\bigr].
\]

The applicant chooses $b$ to maximize this payoff. Consequently, an implementable completion strategy must satisfy, for every $\theta$, $\varepsilon$, and offered requirement $r$,
\begin{equation}
    b_\theta(\varepsilon;r)
    \bigl[v-K_\theta(r)-\varepsilon\bigr]
    \ge
    \widehat b
    \bigl[v-K_\theta(r)-\varepsilon\bigr]
    \qquad
    \text{for all }\widehat b\in\{0,1\}.
    \tag{IC}
    \label{eq:app_screening_ic}
\end{equation}

These are the incentive constraints of the indirect mechanism: the prescribed completion decision must be optimal relative to both available actions. Equivalently,
\[
\begin{aligned}
    b_\theta(\varepsilon;r)=1
    &\quad\Longrightarrow\quad
    v-K_\theta(r)-\varepsilon\ge 0,\\
    b_\theta(\varepsilon;r)=0
    &\quad\Longrightarrow\quad
    v-K_\theta(r)-\varepsilon\le 0.
\end{aligned}
\]

The first condition prevents an applicant from being induced to complete an unprofitable requirement. The second prevents a prescribed withdrawal when completion would be strictly profitable.
Because withdrawal gives the applicant zero, individual rationality requires
\begin{equation}
    b_\theta(\varepsilon;r)
    \bigl[v-K_\theta(r)-\varepsilon\bigr]
    \ge 0.
    \tag{IR}
    \label{eq:app_screening_ir}
\end{equation}

Constraint~\eqref{eq:app_screening_ir} is already implied by \eqref{eq:app_screening_ic}, taking $\widehat b=0$. Individual rationality therefore does not require every applicant to complete recourse. Applicants whose realized costs exceed $v$ withdraw and obtain their outside option. Outright acceptance and rejection also satisfy individual rationality, giving payoffs $v>0$ and $0$, respectively.

The applicant need not observe the signal to make this decision. Once $\theta$, $r$, and $\varepsilon$ are known, both the cost of completion and its reward are known. Beliefs about the principal's signal do not enter the comparison between completion and withdrawal.

\paragraph{The principal's constrained recourse problem}
For a candidate measurable completion strategy $b=(b_H,b_L)$, define
\[
    \widetilde q_\theta(r;b)
    \equiv
    \frac{1}{\bar\varepsilon}
    \int_0^{\bar\varepsilon}
        b_\theta(\varepsilon;r)\,d\varepsilon.
\]

This is the probability that a type-$\theta$ applicant completes the requirement under $b$. At posterior $p$, the principal's best implementable recourse offer solves
\begin{equation}
\begin{aligned}
    V^O(p)
    =
    \max_{r,b_H,b_L}\quad&
    p\,\widetilde q_H(r;b)(a_H+\delta r)
    \\
    &+
    (1-p)\,\widetilde q_L(r;b)(-a_L+\delta r)
    \\
    \text{subject to}\quad&
    r\in\mathcal R,
    \\
    &
    b_\theta(\cdot;r):
    [0,\bar\varepsilon]\longrightarrow\{0,1\}
    \text{ measurable},
    \\
    &
    \eqref{eq:app_screening_ic}
    \text{ and }
    \eqref{eq:app_screening_ir}
    \text{ for each }\theta
    \text{ and }\varepsilon.
\end{aligned}
\label{eq:app_screening_program}
\end{equation}

Writing the problem as a choice over $r$ and $b$ does not give the principal direct control over completion. Rather, the incentive constraints restrict $b$ to behavior that the chosen requirement can induce.

The objective includes the productivity gain $\delta r$ only when recourse is completed. The applicant bears the cost $K_\theta(r)+\varepsilon$; this cost is not a transfer received by the principal. Rejection following withdrawal gives both players zero.

\paragraph{Reduction to completion probabilities}
Constraint~\eqref{eq:app_screening_ic} determines the completion strategy:
\[
    b_\theta^*(\varepsilon;r)
    =
    \mathbf{1}\bigl\{
        \varepsilon\le v-K_\theta(r)
    \bigr\}.
\]

We select completion at indifference, as in the main text. Because the cost shock is continuously distributed, this selection does not affect expected payoffs. Integrating over $\varepsilon$ gives
\begin{equation}
    q_\theta(r)
    =
    \min\left\{
        \max\left\{
            \frac{v-K_\theta(r)}{\bar\varepsilon},
            0
        \right\},
        1
    \right\}.
    \label{eq:app_screening_completion}
\end{equation}

Thus the completion probabilities are induced by the applicant's incentive constraints, not chosen independently by the principal.
Substituting~\eqref{eq:app_screening_completion} into \eqref{eq:app_screening_program} yields
\[
\begin{aligned}
    V^O(p)
    &=
    \max_{r\in\mathcal R} U^O(p,r),\\
    U^O(p,r)
    &=
    p q_H(r)(a_H+\delta r)
    +(1-p)q_L(r)(-a_L+\delta r).
\end{aligned}
\]

This is the recourse problem studied in the main text. Likewise, the applicant's expected payoff before the cost shock is drawn is
\[
    B_\theta(r)
    =
    \mathbb E_\varepsilon
    \left[
        \max\{v-K_\theta(r)-\varepsilon,0\}
    \right],
\]
which incorporates his option to withdraw after observing the shock.
\paragraph{The principal's ex ante policy problem}
Let
\[
    g(s)=\pi f_H(s)+(1-\pi)f_L(s)
\]
denote the unconditional density of the signal, and let $\Phi$ be the set of feasible measurable decision rules. After imposing the applicant's incentive constraints, define the principal's conditional payoff from action $d\in\mathcal A$ by
\[
    \mathcal U(p,d)
    =
    \begin{cases}
        U^A(p)=p a_H-(1-p)a_L,
        & d=A,\\[2pt]
        0,
        & d=R,\\[2pt]
        U^O(p,r),
        & d=(O,r).
    \end{cases}
\]

The principal chooses the decision rule to maximize ex ante expected payoff:
\begin{equation}
    \Pi^*
    =
    \max_{\varphi\in\Phi}
    \int_S
\mathcal U\bigl(p(s),\varphi(p(s))\bigr)
        g(s)\,ds.
    \label{eq:app_screening_exante}
\end{equation}

There are no capacity or budget constraints linking decisions at different signals. Nor are there incentive constraints allowing the applicant to select another signal's offer: the signal is observed by the principal, not reported or chosen by the applicant. The objective therefore separates across signal realizations. Defining
\[
    V(p)
    \equiv
    \max\bigl\{U^A(p),0,V^O(p)\bigr\},
\]
the optimal ex ante payoff is
\[
    \Pi^*
    =
    \int_S V\bigl(p(s)\bigr)g(s)\,ds.
\]

An optimal decision rule selects a maximizing action at each posterior, with ties resolved in favor of acceptance over recourse and recourse over rejection. Compactness of $\mathcal R$ and continuity of the payoff functions permit a measurable selection. The posterior-by-posterior maximization in the main text therefore solves the principal's ex ante mechanism-choice problem within the specified contract space.

\paragraph{The nature of screening}
The cost ordering $K_H(r)\le K_L(r)$ implies $q_H(r)\ge q_L(r)$. Hence profitable applicants are weakly more likely to complete any given requirement. Whenever completion has positive probability, the principal's posterior after observing completion is
\[
    \Pr(\theta=H\mid s,\text{ completion})
    =
    \frac{p(s)q_H(r)}
    {p(s)q_H(r)+(1-p(s))q_L(r)}
    \ge p(s).
\]

The inequality is strict when $0<p(s)<1$ and $q_H(r)>q_L(r)$. Screening need not produce perfect separation: both types may complete with positive probability. Recourse can improve the composition of accepted applicants while also raising the principal's payoff from each completer by $\delta r$.

There are no additional truth-telling constraints comparing contracts assigned to reported types $H$ and $L$, because the mechanism does not solicit a type report. At a given score, both types face the same offer and choose between completion and withdrawal. Requirements may differ across scores because scores are observed by the principal, not because applicants select from a menu.

Moreover, within the maintained contract class, a menu containing only different recourse requirements would not induce applicants to select distinct positive requirements. Since $K_\theta'(r)>0$, every applicant who would complete a more demanding requirement would strictly prefer to complete a less demanding requirement that delivers the same acceptance benefit. The relevant self-selection margin is therefore completion versus withdrawal, rather than choice among requirements. Allowing transfers or report-contingent acceptance lotteries would enlarge the mechanism space and introduce additional incentive constraints. The optimality results in this paper concern the deterministic, no-transfer mechanisms defined above.

\section{Research on algorithmic recourse in computer science}

\begin{figure}[h]
    \centering
    \includegraphics[width=0.95\textwidth]{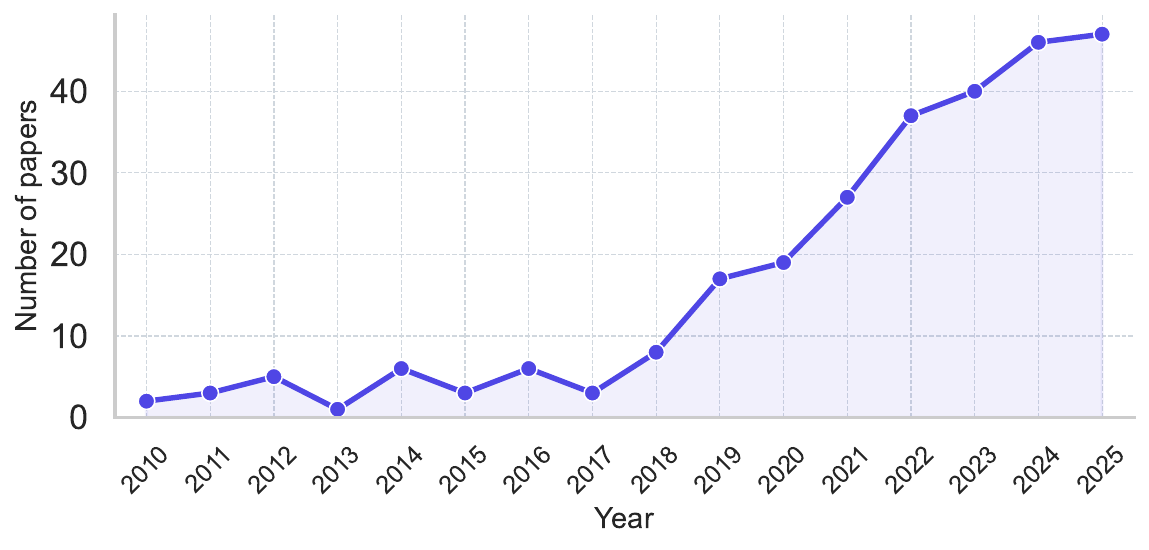}
    \caption{ArXiv papers mentioning ``algorithmic recourse''}
    \label{fig:arxiv_trend}
\end{figure}

A growing computer science literature on algorithmic recourse focuses on designing algorithms to generate recourse steps that can be introduced into decision systems. Figure~\ref{fig:arxiv_trend} illustrates the rapid expansion of this research area, showing growth in the number of papers on arXiv including the term ``algorithmic recourse'' in recent years. 

\end{document}